\documentclass[aps,prc,reprint,superscriptaddress,showpacs,nofootinbib]{revtex4-1}

\usepackage{color}
\usepackage{framed}{
\usepackage{amsmath}
\usepackage{amssymb}
\usepackage{graphicx}
\usepackage[normalem]{ulem}

\def\fic{\text{fic}}

\begin{document}

\title{
Elementarity of composite systems
} 

\author{Hideko Nagahiro}
\thanks{}
\affiliation{Department of Physics, Nara Women's University, 
Nara 630-8506, Japan}
\affiliation{Research Center for Nuclear Physics (RCNP), Osaka
University, Ibaraki, Osaka 567-0047, Japan}

\author{Atsushi Hosaka}
\affiliation{Research Center for Nuclear Physics (RCNP), Osaka
University, Ibaraki, Osaka 567-0047, Japan}
\affiliation{J-PARC Branch, KEK Theory Center, 
KEK, Tokai, Ibaraki 319-1106, Japan}

\date{\today}

\begin{abstract}
The ``compositeness'' or ``elementarity'' is investigated for
 $s$-wave composite states dynamically generated by energy-dependent
 and independent interactions.  The bare mass of the corresponding
 fictitious elementary particle in an equivalent Yukawa model is shown
 to be infinite, {indicating} that the wave function renormalization
 constant $Z$ is equal to zero.  The idea can be equally {applied to} both
resonant and bound states.  In a special case
 {of} zero-energy bound states, the condition $Z=0$ does not necessarily
 mean that the elementary particle has the infinite bare mass.  {We
 {also} emphasize}
 arbitrariness {in} the ``elementarity'' {leading} to multiple
 interpretations of a physical state, which can be {either}
 a pure composite {state} with $Z=0$ {or} an elementary particle with 
 $Z\ne 0$.  {The arbitrariness is unavoidable because the
 renormalization constant $Z$ is not a
 physical observable.}
\end{abstract}


\maketitle

%
\section{Introduction}
Observations of the new hadrons have been indicating the
existence of the so-called exotic
hadrons~\cite{Choi:2003ue,Aubert:2004ns,Choi:2007wga,Aaij:2013zoa}.
Since many of them have been found in the threshold region, they are
expected to develop a structure of hadronic composite; a
loosely bound or resonant system of constituent
hadrons.
Recently, the hadronic composite states have been studied
extensively in the context of the dynamically generated
states~\cite{Oller:1998hw,Baru:2003qq,Jido:2003cb,PhysRevC.65.035204,Roca:2005nm,Hyodo:2008xr},
{while there are also approaches which take into account the effect
of hadron dynamics for $\bar{q}q$
mesons~\cite{Black:2000qq,Urban:2001ru,Fariborz:2005gr,Fariborz:2006ff,Parganlija:2012fy}.} 
Since the scales of composite and elementary states are not well
separated, one would naturally ask how ``composite'' the composite state{s}
are, or to what extent the composite states contain ``elementary''
components.  {The similar issue for
the relation between the composite state and elementary state has been
discussed in Ref.~\cite{Giacosa:2009bj}}.

{The question of ``compositeness'' or ``elementarity'' has been
studied from as early as 1960's by using the wave function
renormalization constant
$Z$~\cite{Weinberg:1962hj,Weinberg:1965zz,Lurie:1964,*Lurie:book}.
Employing a four-point fermi model for a 
composite state and a Yukawa model for an elementary particle, it was
shown}
that the wave function renormalization constant $Z$
for a bound state
should be equal to
zero, which is {the} so-called the compositeness condition
$Z=0$~\cite{Lurie:1964,*Lurie:book}.   
The attempts have been made not only for bound states but also for
resonant states
recently~\cite{PhysRevC.85.015201,Nagahiro:2011jn,Nagahiro:2013hba},
{although the meaning of the renormalization constant $Z$ for
resonant states is still controversial.}

Here in this article, we show that the wave function renormalization
constant $Z$ can be zero for any composite state dynamically generated
by an energy-dependent interaction like the Weinberg-Tomozawa term,
as {in the case for} {a bound state by} an energy-independent
interaction~\cite{Lurie:1964,*Lurie:book}. 
{We show that it is possible to construct a}
Yukawa model
which {gives the completely equivalent scattering amplitude to the one
obtained by the Weinberg-Tomozawa type interaction, by letting}
the bare mass and {the} bare coupling of the fictitious
elementary particle infinite. 
{The above idea can be applied not only to bound states but also to
resonances, although the essential concept was pointed out by Weinberg
in Ref.~\cite{Weinberg:1962hj}.}

At the same time, we investigate a difficult{y} of the measurement
of the elementarity by means of $Z$ {due to model-dependence
of the renormalization constant}.  We show that
multiple interpretations 
of the physical state are possible, and the elementarity cannot
be evaluated {from experiments} in a model-independent manner.  
We {emphasize} that only {by} specifying a
 model with a definite cut-off scale,
 we can make a meaningful measurement by the constant $Z$. 

We also discuss that the underlying mechanism of $Z=0$ for the
zero-energy bound state can be different from that of finite binding
energy or resonant states.  We show that the condition $Z=0$ for a
{barely bound system, like the deuteron,}
does not necessarily mean that
{the} {corresponding} elementary particle has the 
infinite bare mass, {and does not {exclude}
an elementary state (such as a quark-core {of} $qqq$ {for
baryons} or $q\bar{q}$ {for mesons})
close to the physical state.}

{In this article, most of the discussions are made for mesons, however
the results can be also applied to baryons.}
This article is organized as follows.  In Sec.~\ref{sec:compositenessZ}
we {show how in a Yukawa model we can introduce the elementary
particle which is equivalent}
to the $s$-wave state dynamically generated, and show its
wave function renormalization constant $Z$ is zero.  In
Sec.~\ref{sec:multiple_int}, we investigate an arbitrariness of $Z$
which leads to multiple interpretation of a physical state, {by taking
the sigma ($\sigma(500)$) resonance in the sigma model} as
an example.  In Sec.~\ref{sec:BE=0} we discuss the {unique feature
of} the zero-energy bound
state.  Finally, Sec.~\ref{sec:conclusion} 
is devoted to the summaries {and discussions}.

\section{The condition $Z=0$ for composite states}
\label{sec:compositenessZ}
\subsection{A brief review of Lurie's discussion; {energy
  independent interaction}}
\label{sec:Lurie}
{We start from a brief review of the ``compositeness condition
$Z=0$'' discussed in Ref.~\cite{Lurie:1964,*Lurie:book}.}
{There}, authors {compared} 
the four-fermi {model} and a Yukawa model, {and studied their
equivalence in terms of their scattering amplitudes.}
{Here, we revisit the compositeness condition {for} a meson-meson
bound system.}  

\begin{figure}[t]
\includegraphics[width=0.4\textwidth]{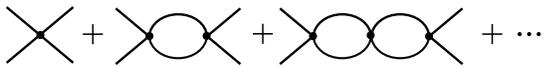}
\caption{Sum of the infinite set of diagrams that contributes to the 
 {meson-meson} scattering amplitude (\ref{eq:tf}).   
\label{fig:WT_pole}}
\end{figure}

{Let us consider} a four-point interaction with a constant coupling
$v$.  
The {meson-meson} scattering amplitude {$t(s)$} is obtained by {summing up} the
infinite set of diagrams as shown  
in Fig.~\ref{fig:WT_pole},
\begin{equation}
t(s)=v + v G(s) v + \cdots = \frac{1}{v^{-1}-G(s)} \ ,
\label{eq:tf}
\end{equation}
{where $G(s)$ denotes the integrated two-body {bosonic} propagator given as a
function of the total energy {square} $s$ of the system {as}
\begin{equation}
G(s)=i\int\frac{d^4q}{(2\pi)^4}\frac{1}{q^2-m_1^2+i\epsilon}
\frac{1}{(P-q)^2-m_2^2+i\epsilon} \ .
\label{eq:G}
\end{equation}
Here $P=(\sqrt{s},0,0,0)$, and $m_1$ and $m_2$ are the masses of the
{two mesons}.
{We regularize} the loop function $G(s)$ 
{appropriately} by using, for example, the dimensional
regularization, the three-dimensional cut-off scheme, and so on.}
If the interaction $v$ is attractive enough, the amplitude develops a
pole {at} $\mu^2$ satisfying $ v^{-1} - G(\mu^2)=0$ as a bound state {of
the two mesons.}
The loop function $G(s)$ is expanded as a Taylor series about $\mu^2$
\begin{equation*}
 G(s) = G(\mu^2) + (s-\mu^2)G'(\mu^2) + (s-\mu^2)G_h(s)
\end{equation*}
where $G_h(s)$ contains higher order terms and {vanishes} at $\mu^2$,
{$G_h(\mu^2)=0$}.
The scattering amplitude is then given by
\begin{equation}
 t(s) = g^2(s)\frac{1}{s-\mu^2} ,
\label{eq:t_F}
\end{equation}
where the coupling $g(s)$ of the bound state to the two mesons is
defined by 
\begin{equation}
g^2(s)=-\frac{1}{G'(\mu^2)+G_h(s)} \ .
\label{eq:g2} 
\end{equation}
{Hereafter we refer to a model which generates such a composite state
as a {\em composite model}.}

\begin{figure}[hbt]
\includegraphics[width=0.4\textwidth]{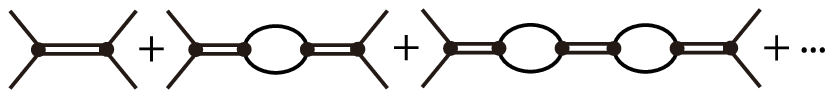}
\caption{Scattering amplitude in the Yukawa model.
\label{fig:Yukawa}}
\end{figure}

Now, we consider a {\em Yukawa model} which 
has only a three-point interaction of an elementary particle {with
the two mesons}.  With a {bare} coupling constant $g_0$, the
{full} scattering amplitude as 
shown in Fig.~\ref{fig:Yukawa} is given by
\begin{equation}
 t_Y(s)=g_0^2 \Delta(s) \ ,
\label{eq:t_Y}
\end{equation}
where $\Delta(s)$ is the dressed propagator given by
\begin{equation}
 \Delta(s)=\frac{1}{s-M_0^2-g_0^2G(s)}  \ .
\end{equation}
{Here we have assumed that the loop function $G(s)$ in the Yukawa
model is the same as that in the composite model.}
If the amplitudes $t_Y$ {of} Eq.~(\ref{eq:t_Y}) in the Yukawa model
and $t$ {of} Eq.~(\ref{eq:t_F}) in the composite model are equal, $t_Y$ should have
a pole at the same position $\mu^2$ as $t$.  By expanding the loop function
$G(s)$ again, we {obtain}
\begin{equation}
 t_Y=\frac{g_0^2}{1-g_0^2(G'(\mu^2)+G_h(s))}\frac{1}{s-\mu^2} \ ,
\end{equation}
from which the renormalized coupling $g_R$ at the pole
position is defined by
\begin{equation}
 g_R^2=Zg_0^2 \ .
\label{eq:gR2=Zg02}
\end{equation}
{Here} $Z$ is the wave function renormalization {constant
defined by}
\begin{eqnarray}
 Z&=&\left(1-g_0^2G'(\mu^2)\right)^{-1} \label{eq:Z_const0}\\
&=&1+g_R^2G'(\mu^2) \ .
\label{eq:Z_const}
\end{eqnarray}
Equivalence of $t$ and $t_Y$ requires that the
renormalized coupling $g_R^2$ at the pole position is equal to $g^2$
in Eq.~(\ref{eq:g2})
\begin{equation}
 g_R^2(\mu^2) {=g^2(\mu^2)}= -\frac{1}{G'(\mu^2)} \ .
\label{eq:GR}
\end{equation}
With Eqs.~(\ref{eq:Z_const}) and (\ref{eq:GR}), one can
conclude that the wave function renormalization {constant} for {the bound state}
is zero.
It implies that the
bare, unrenormalized, field $\phi=Z^{1/2}\phi_R$ in the Yukawa model
vanishes for the composite boson.  
This is the content of the so-called ``compositeness condition
$Z=0$'' from the field theoretical point of view discussed in
Ref.~\cite{Lurie:1964,*Lurie:book}.  

The relation
(\ref{eq:Z_const}) between the wave function renormalization
constant $Z$ and the
renormalized coupling $g_R$ is employed also in the
estimation of the compositeness for the deuteron system in
Ref.~\cite{Weinberg:1965zz}.
{There} the (renormalized) coupling $g_R$ is estimated by
experimental data of the low-energy
$p$-$n$ scattering. 
In
Ref.~\cite{Weinberg:1965zz}, the $p$-$n$-$d$ coupling does not have an
energy-dependence, and 
the non-relativistic form of the loop function is employed.
The estimated $Z$ in Ref.~\cite{Weinberg:1965zz}
corresponds to the wave function renormalization constant for the
{fictitious} elementary particle (deuteron) in the Yukawa
model~(\ref{eq:t_Y}) with the constant coupling.

The above discussion cannot be directly applied {to} a
resonant state.
One way to allow an
$s$-wave resonance is to take the interaction $v$
energy dependent.  
{It turns out}
that the scattering amplitude with an
energy-dependent $v(s)$ cannot be replaced by the Yukawa amplitude $t_Y$
in Eq.~(\ref{eq:t_Y}) with the energy-independent coupling $g_0$.
{Instead,}
we introduce an
equivalent Yukawa model to a composite model with the
energy-dependent interaction $v(s)$ where the both models
give completely the same
scattering amplitude, and discuss the wave function renormalization constant.

\subsection{Energy dependent interaction; Weinberg-Tomozawa type}
\label{sec:WT}
{Let us} consider a composite model with an energy-dependent
interaction $v(s)$ which generates dynamically an $s$-wave composite
state.
{In this section, we consider a specific form for the energy-dependent
interaction, that is the Weinberg-Tomozawa (WT) type, and then later we
generalize it in Sec.~\ref{sec:gene}.
The WT interaction takes the following form,}
\begin{equation}
 v(s)=-\frac{1}{f^2}(s-m^2) \ ,
\label{eq:v1}
\end{equation}
{wh}ere $f$ and $m$ are constants
{having dimensions of {mass}}.  This {is a} contact
interaction as shown in the first diagram in Fig.~\ref{fig:WT_pole}.

By summing up the infinite set of diagrams as shown in
Fig.~\ref{fig:WT_pole}, we obtain the  
{scattering} amplitude $T$ as
\begin{equation}
 T(s)=\frac{1}{v(s)^{-1}-G(s)} \ ,
\label{eq:t1}
\end{equation} 
{where the on-shell factorization is employed~\cite{Oller:1997ti}.
Here $G(s)$ is the loop
function in Eq.~(\ref{eq:G}) which is regularized appropriately.}
If the potential $v(s)$ is attractive enough, the amplitude develops a
pole at mass $\mu$ satisfying
\begin{equation}
 v(\mu^2)^{-1}-G(\mu^2)=0 \ ,
\label{eq:pole1}
\end{equation}
{with the regularized $G$.}
The pole corresponds to a bound state if {it} appears below the
threshold, or to a resonant state if above the threshold.  {The resonant
state is a consequence of the energy dependent interaction $v(s)$, while
a constant interaction $v$ {can generate only a bound state}.}

{Before} constructing a Yukawa model giving the scattering amplitude 
{which is exactly equivalent to} $T(s)$ in Eq.~(\ref{eq:t1}), we {attempt to} shift
the denominator of Eq.~(\ref{eq:t1}) by a constant $\delta (>0)$ as
\begin{equation}
 T_\delta(s)=\frac{1}{v(s)^{-1} - G(s) + \delta} \ ,
\label{eq:Ta}
\end{equation}
{and later {let} $\delta$ {be} zero again.}
{It}
is clear that the shifted amplitude $T_\delta(s)$ smoothly 
{reduces to} the original $T(s)$ in the limit $\delta \rightarrow
0$ as
\begin{equation*}
 \lim_{\delta \rightarrow 0} T_\delta(s)=T(s) \ .
\end{equation*}
The shifted amplitude $T_\delta(s)$ has a pole at $\mu_{\delta}^2$
satisfying 
\begin{equation}
 v(\mu^2_{\delta})^{-1} - G(\mu^2_\delta) + \delta =0 \ ,
\label{eq:mu_a}
\end{equation}
which {also reduces to} $\mu^2$ in the limit $\delta\rightarrow 0$ as
\begin{equation*}
 \lim_{\delta \rightarrow 0} \mu_{\delta}^2 = \mu^2 \ .
\end{equation*}
The {inverse of the} interaction kernel $v(s)^{-1}$ is expanded as a
Taylor series about 
the pole $\mu_\delta^2$ as
\begin{eqnarray}
 v(s)^{-1} &=& - \frac{f^2}{s-m^2} \nonumber \\
&=& - \frac{f^2}{\mu_\delta^2-m^2} + \frac{f^2(s-\mu_\delta^2)}{(\mu_\delta^2-m^2)^2} 
+ (s-\mu_\delta^2)v_h(s) \nonumber \\
\label{eq:vinv}
\end{eqnarray}
where $v_h(s)$ {contains} {higher order} terms and becomes zero at
$s=\mu_\delta^2$.  Similarly the function $G(s)$ is expanded {about $\mu_\delta^2$} as
\begin{equation}
 G(s) = G(\mu_\delta^2) + (s-\mu_\delta^2)G'(\mu_\delta^2) + (s-\mu_\delta^2)G_h(s) \ .
\label{eq:expG}
\end{equation}
By using Eqs.~(\ref{eq:mu_a}), (\ref{eq:vinv}) and (\ref{eq:expG})
the amplitude $T_\delta(s)$ {can be equivalently written as}
\begin{equation}
 T_\delta(s)= g_R^2(s)\frac{1}{s-\mu_\delta^2} \ ,
\end{equation}
where
\begin{equation}
g^2_R(s)=\left(\frac{f^2}{(\mu_\delta^2-m^2)^2}
+v_h(s) - G'(\mu^2_\delta) - G_h(s)\right)^{-1}
\label{eq:gR}
\end{equation}
{is interpreted as an} effective coupling of the composite state {to
the two mesons}.  At the pole position $s=\mu_\delta^2$ {it is
given by,}
\begin{equation}
g^2_R(\mu_\delta^2)=\left(\frac{f^2}{(\mu_\delta^2-m^2)^2}
- G'(\mu^2_\delta)\right)^{-1} \ .
\label{eq:gR1}
\end{equation}

Now let us {construct a} Yukawa {model} giving exactly the
same {scattering} amplitude {in the composite model in} 
Eq.~(\ref{eq:t1}).  The prescription is similar to the one developed in
Ref.~\cite{Hyodo:2008xr} in which the CDD-pole component is discussed.
Let us define a function $V_Y(s)$ by
\begin{equation}
 V_Y(s)^{-1} \equiv v(s)^{-1} + \delta
\label{eq:V_Y}
\end{equation}
so that
\begin{equation}
 V_Y(s)=\frac{\frac{1}{\delta}(s-m^2)}{s-m^2-\frac{f^2}{\delta}} .
\end{equation}
{By} defin{ing} $m_\fic^2$ by
\begin{equation}
 m_\fic^2 \equiv \frac{f^2}{\delta} + m^2 \ 
\label{eq:mfic}
\end{equation}
{and eliminating $\delta$, we find an expression}
\begin{equation}
 V_Y(s)=\frac{1}{f^2}(m_\fic^2
  -m^2)(s-m^2)
\frac{1}{s-m_\fic^2} \ .
\label{eq:V_Y1}
\end{equation}
Here
$m_\fic$ can be interpreted as {the} bare mass of the
fictitious elementary particle.
The function in front of the propagator
is interpreted as {the} bare coupling of the fictitious
particle to {the two mesons} defined as
\begin{equation}
 g_0^2(s;m_\fic) \equiv \frac{1}{f^2}(m_\fic^2
  -m^2)(s-m^2) \ . \label{eq:g01}
\end{equation} 
With these interpretations we can regard $V_Y(s)$ as a Yukawa pole term
with the energy-dependent coupling $g_0(s;m_\fic)$.
{We note that, in
Ref.~\cite{Hyodo:2008xr}, an additional CDD-pole {term} is
defined by subtracting the WT {term} $v(s)$ from 
$V_Y(s)$ in Eq.~(\ref{eq:V_Y1}). Here the trick in {our} study is to regard
the whole {of} $V_Y(s)$ {of} Eq.~(\ref{eq:V_Y1}) as the 
Yukawa pole
term, and the constant $\delta$ is treated just as a parameter}.

Now we {shall see that} the scattering amplitude {of the
composite model} in Eq.~(\ref{eq:t1}) {can}
be generated by the Yukawa {model} as shown in Fig.~\ref{fig:Yukawa} as
\begin{equation}
 T_Y(s) = g_0^2(s;m_\fic) {\Delta(s)} \ ,
\label{eq:T_Y}
\end{equation}
{where} the {dressed} propagator {$\Delta(s)$} for the fictitious
elementary particle is given by 
\begin{equation}
 \Delta(s)=\frac{1}{s-m_\fic^2-g_0^2(s;m_\fic) G(s)} \ .
\label{eq:Yukawa1}
\end{equation}
{The} scattering amplitude {(\ref{eq:T_Y})} is identical with the shifted amplitude
$T_\delta(s)$ in Eq.~({\ref{eq:Ta}})
and then {reduces} to $T(s)$ in Eq.~(\ref{eq:t1}) in the limit 
$\delta\rightarrow 0$,
\begin{equation*}
 \lim_{\delta\rightarrow 0} T_Y(s) = T(s) \ .
\end{equation*}
In this {manner}, the {Yukawa} pole {term $V_Y(s)$} 
of Eq.~(\ref{eq:V_Y1}) {in the Yukawa model}
is equivalent to the four-point {WT} type
interaction of Eq.~(\ref{eq:v1}) {{in} the composite model}.

Since we {have} define{d} the {dressed} propagator $\Delta(s)$
of the fictitious {elementary} Yukawa 
particle {as} in Eq.~(\ref{eq:Yukawa1}), we can {evaluate} the wave function
renormalization {constant} {for it}.   We 
expand the self-energy {defined by}
\begin{equation}
 \Pi(s;m_\fic)\equiv g_0^2(s;m_\fic^2)G(s)
\label{eq:PiY}
\end{equation}
 as a Taylor series
about the pole $\mu^2_{\delta}$ as  
\begin{eqnarray}
\Pi(s;m_\fic) &=& \Pi(\mu^2_{\delta};m_\fic)  + (s-\mu^2_{\delta})
 \Pi'(\mu^2_{\delta};m_\fic) 
 \nonumber \\
&+& (s-\mu^2_{\delta})\Pi_h(s)
\label{eq:Pi}
\end{eqnarray}
where again $\Pi_h$ {contains} {higher order} terms and becomes zero at
$s=\mu^2_\delta$.  {T}he {dressed} propagator {$\Delta(s)$} in 
Eq.~(\ref{eq:Yukawa1}) has the pole {at}
\begin{eqnarray*}
&& m_\fic^2+\Pi(\mu^2_\delta;m_\fic) \nonumber \\
&=& m_\fic^2+\frac{1}{f^2}(m_\fic^2
  -m^2)(\mu_\delta^2-m^2)G(\mu_\delta^2) \nonumber \\
&=&\mu^2_\delta  \ ,
\end{eqnarray*}
{which is} the same 
as {the pole position of} $T_\delta(s)$. {Here we have used
Eqs.~(\ref{eq:mu_a}) and (\ref{eq:mfic}).} 
The {dressed} propagator ${\Delta}(s)$ {can be
rewritten as,}
\begin{equation*}
 \Delta{(s)} = \frac{1}{1-\Pi'(\mu_\delta;m_\fic)-\Pi_h(s)}\frac{1}{s-\mu^2_{\delta}}
\ ,
\end{equation*}
and the wave function renormalization constant for the fictitious particle
is defined by
\begin{equation}
 Z=\left(\left. 1- \Pi'(s;m_\fic)\right|_{s=\mu^2_\delta}\right)^{-1} \ .
\label{eq:ourZ}
\end{equation}
The derivative of the self-energy $\Pi$ is obtained by
\begin{eqnarray}
\left.\Pi'(s;m_\fic)\right|_{s=\mu^2_\delta} &=&
  \frac{1}{f^2}(m_\fic^2-m^2)G(\mu^2_\delta) \nonumber \\
 &+& \frac{1}{f^2}(m_\fic^2-m^2)(\mu^2_\delta-m^2)G'(\mu^2_\delta) \nonumber \\
 &=& \frac{\mu^2_\delta-m_\fic^2}{\mu_\delta^2-m^2} 
 + {g_0^2(\mu_\delta^2;m_\fic^2)}G'(\mu^2_\delta) \ .
  \nonumber \\
\label{eq:Pi'}
\end{eqnarray}
Note that we have the first term of
Eq.~(\ref{eq:Pi'}) in addition to the derivative of the loop function
$G$ {as in Eq.~(\ref{eq:Z_const0})}.  {This} is the consequence
of the energy dependence of the {bare} coupling $g_0^2$ {and is
non-negligible in the present analysis}.   The inverse of the wave
function renormalization constant
$Z^{-1}$ is obtained by 
\begin{eqnarray}
 Z^{-1} &=& \frac{(\mu^2_\delta-m^2)(m_\fic^2-m^2)}{f^2}
\left\{\frac{f^2}{(\mu^2_\delta-m^2)^2}-G'(\mu_\delta^2)\right\}
\nonumber \\ \\
&=& g_0^2(\mu_\delta^2;m_\fic) \frac{1}{g_R^2(\mu_\delta^2)} \ ,
\label{eq:Z}
\end{eqnarray}
where in the last line we use Eq.~(\ref{eq:gR1}) and (\ref{eq:g01}). 

The Yukawa model with the fictitious {elementary} particle in the limit 
$\delta\rightarrow 0$ is {identical} with the {composite model}
in the sense that both models give the {same} scattering
amplitude {in the whole energy range}.  
{Since the loop function $G(s)$ {has been}
regularized, the renormalized coupling $g_R$ remains finite in the limit
$\delta\rightarrow 0$, except for a singular point $G'(s)$ at the
threshold.  (We will return to this point later in Sec.~\ref{sec:BE=0}.)}
Because the mass of the fictitious particle $m_\fic$ {in
Eq.~(\ref{eq:mfic})} {diverges} in the limit 
$\delta\rightarrow 0$, the bare coupling $g_0$ also {does;}
\begin{eqnarray*}
 \lim_{\delta\rightarrow 0} m_\fic^2 = \infty \ , \\
 \lim_{\delta\rightarrow 0} g_0^2 = \infty \ .
\end{eqnarray*}
As a consequence, the wave function renormalization constant $Z$ must be zero in
the limit $\delta \rightarrow 0$ as
\begin{equation*}
 \lim_{\delta\rightarrow 0} Z = 0 \ .
\end{equation*}
This means that any composite state dynamically generated by the
WT type interaction can be represented by a fictitious
elementary Yukawa particle with $Z=0$ whose bare field
$\phi=Z^{1/2}\phi_R$ {vanishes}.  This conclusion does not depend on
cut-off scale.

We stress here that the condition $Z=0$ for the composite {state} is
not due to the divergence of the loop function $G$, {nor $G'$},
{but those of the self-energy $\Pi$ and $\Pi'$} {as implied in
Eq.~(\ref{eq:ourZ}).} 
The underlying mechanism of $Z=0$ is
that the bare coupling
$g_0$ {in the Yukawa model} is proportional to the 
bare mass of the fictitious particle $m_\fic$, {which} diverges
to be consistent with the composite model.
In the present Yukawa model, the fictitious elementary particle
with the infinite mass becomes the physical resonant state {by
the one-loop correction} with the infinite Yukawa coupling $g_0$ and the
finite (regularized) loop function $G(s)$.

{Here, we would like to note that our observation {of $Z=0$} differs from the
argument in Ref.~\cite{PhysRevC.85.015201}, where the wave function
renormalization {constant} for a composite state generated by the WT
interaction is not zero.  This difference comes from the different
definitions of the corresponding Yukawa models.  The wave function
renormalization {constant} $Z$ defined in Ref.~\cite{PhysRevC.85.015201} is for a
Yukawa particle with a {\em constant} coupling as {defined in}
Eq.~(\ref{eq:Z_const})~\cite{Lurie:1964,*Lurie:book,Weinberg:1965zz}.
In contrast, {here} we have shown {that}
{it is possible to construct} the Yukawa model which is {\em completely
equivalent} with the composite model with the energy dependent WT
interaction {by allowing} an energy-dependent coupling $g_0(s)$ (\ref{eq:g01}). 

{To see {a role of the energy-dependence of $g_0$ more}
clearly, we rewrite $Z$ in the 
energy-dependent Yukawa model in Eq.~(\ref{eq:ourZ}) as,
\begin{equation}
 Z=\frac{1}{1-\frac{\mu^2_\delta-m_\fic^2}{\mu^2_\delta-m^2}}
\left(1+g_R^2(\mu^2_\delta)G'(\mu^2_\delta)\right)
\label{eq:ourZ2}
\end{equation}
where we {have} used Eqs.~(\ref{eq:ourZ}), (\ref{eq:Pi'}) and
(\ref{eq:Z}).  By 
comparing Eqs.~(\ref{eq:Z_const}) and (\ref{eq:ourZ2}), we can see that
the difference is the term $(\mu^2_\delta-m_\fic^2)/(\mu^2_\delta-m^2)$,
which comes from the first term of Eq.~(\ref{eq:Pi'}) due to the energy
dependence of the Yukawa coupling.  Equation (\ref{eq:ourZ2}) can be
{further} rewritten as
\begin{eqnarray*}
 Z &=& \frac{\mu^2_\delta-m^2}{m_\fic^2-m^2}
\left(1+g_R^2(\mu^2_\delta)G'(\mu^2_\delta)\right) \\
&=& \frac{\delta}{f^2}(\mu_\delta^2-m^2)\left(1+g_R^2(\mu^2_\delta)G'(\mu^2_\delta)\right) 
\end{eqnarray*}
which becomes zero with $m_\fic^2\rightarrow \infty$ or
$\delta\rightarrow 0$.
Unlike the case {of} the constant interaction $v$, with the WT
interaction $v(s)$ the renormalized coupling $g_R^2$ is not equal to
$-1/G'$.  Instead, the additional term
$(\mu^2_\delta-m_\fic^2)/(\mu^2_\delta-m^2)$ owing to the energy
dependence of the Yukawa model plays {the crucial role to achieve}
$Z=0$.
}

\subsection{Energy dependent interaction; general form}

\label{sec:gene}
\label{sec:general_v}
In the previous section, we have {considered} the WT type
interaction to generate {a} composite state.   Here we generalize the
discussion {to} a general form of the interaction kernel and discuss the
{requirement to obtain} $Z=0$.  
{Let us assume that the interaction kernel $v(s)$ has an
energy-dependence and} {is} attractive enough to generate a resonant or
bound state.
We again start from the shifted amplitude
$T_\delta(s)=(v^{-1}(s)-G(s)+\delta)^{-1}$ by a constant $\delta(>0)$.
Following the same procedure in Eqs.~(\ref{eq:vinv})--(\ref{eq:gR1}),
the (renormalized) coupling $g_R$ is obtained {at $\mu_\delta^2$} as
\begin{equation}
g_R^2(\mu_\delta^2)=
\frac{v(\mu_\delta^2)}{v'(\mu_\delta^2)(\delta-G(\mu_\delta^2))-v(\mu_\delta^2)G'(\mu^2_\delta)} \ .
\end{equation}
The equivalent Yukawa pole term is {constructed from}
\begin{equation}
 V_Y(s) = \frac{1}{v(s)^{-1}+\delta} = \frac{v(s)}{1+v(s)\delta} \ .
\end{equation}
For {an} attractive $v(s)$, $V_Y(s)$ {has a pole}
for a positive $\delta$ at the energy satisfying  
\begin{equation}
v(s)=-\frac{1}{\delta} \ .
\label{eq:m_fic}
\end{equation}
The mass of the fictitious {elementary} particle $m_\fic$ is defined by the
solution of Eq.~(\ref{eq:m_fic}).
We expand $v(s)$ about $m_\fic^2$ as
\begin{equation*}
 v(s)=v(m_\fic^2)+(s-m_\fic^2)h(s;m_\fic^2)
\end{equation*}
and rewrite $V_Y$ as an explicit {pole term} as
\begin{equation}
 V_Y(s)=\frac{1}{\delta}\frac{v(s)}{h(s;m_\fic^2)}\frac{1}{s-m_\fic^2} \ .
\end{equation}
By defining the bare coupling $g_0^2$ as
\begin{equation}
 g_0^2(s;m_\fic)=\frac{1}{\delta}\frac{v(s)}{h(s;m_\fic^2)} \ ,
\label{eq:g02}
\end{equation}
the {scattering} amplitude in the Yukawa model is {expressed as}
$$T_Y(s)=g_0^2(s;m_\fic^2)\Delta(s) \ ,$$
{where} the {dressed} propagator of the fictitious particle is given by
$$
\Delta(s)=\frac{1}{s-m_\fic^2-g_0^2(s;m_\fic^2)G(s)} \ .
$$
Since the propagator {$\Delta(s)$} has the pole at $\mu_\delta^2$
satisfying
\begin{equation*}
 m_\fic^2 + g_0^2(\mu_\delta^2;m_\fic)G(\mu_\delta^2) = \mu_\delta^2,
\end{equation*}
the bare coupling $g_0^2$ in Eq.~(\ref{eq:g02}) {can be} expressed
at the pole position as  
\begin{equation}
 g_0^2(\mu_\delta^2;m_\fic) = \frac{\mu_\delta^2-m_\fic^2}{G(\mu_\delta^2)}
\label{eq:g0_gene}
\end{equation}
{W}e
follow Eqs.~(\ref{eq:Pi})--(\ref{eq:Z}) and obtain
\begin{eqnarray}
Z^{-1}&=&\frac{\mu^2_\delta-m_\fic^2}{G(\mu_\delta^2)v(\mu_\delta^2)}
\left\{v'(\mu_\delta^2)(\delta-G(\mu_\delta^2))-v(\mu_\delta^2)G'(\mu^2)
\right\} \nonumber \\
&=&g_0^2(\mu_\delta^2;m_\fic)\frac{1}{g_R^2(\mu^2_\delta)} \ .
\label{eq:Z_gen}
\end{eqnarray}
The renormalized coupling $g_R^2$ is finite in the limit
$\delta\rightarrow 0$ {with the regularized $G(s)$ and $G'(s)$.}
The {requirement to obtain}
$Z=0$ is {again} that $g_0^2$ diverges with
$\delta\rightarrow 0$, {and then}
$m_\fic^2$ diverges {as the solution of Eq.~(\ref{eq:m_fic}).} 

{Now we observe from Eq.~(\ref{eq:m_fic}) that $|v(s)|$ diverges at
$s=m_\fic^2$ $(\delta\rightarrow 0)$.}
As shown in Fig.~\ref{fig:v}(a) (red solid line), the WT interaction
in Eq.~(\ref{eq:v1}) diverges for large $s$, then $m_\fic^2$ is
infinite,
{and therefore} the wave function renormalization {constant} is zero $Z=0$. 
In the light of these discussions, we can expect that an attractive
interaction with a simple polynomial function of $\sqrt{s}$ which
negatively diverge{s} for large $s$ gives $Z=0$.

\begin{figure}[hbt]
\includegraphics[width=0.4\textwidth]{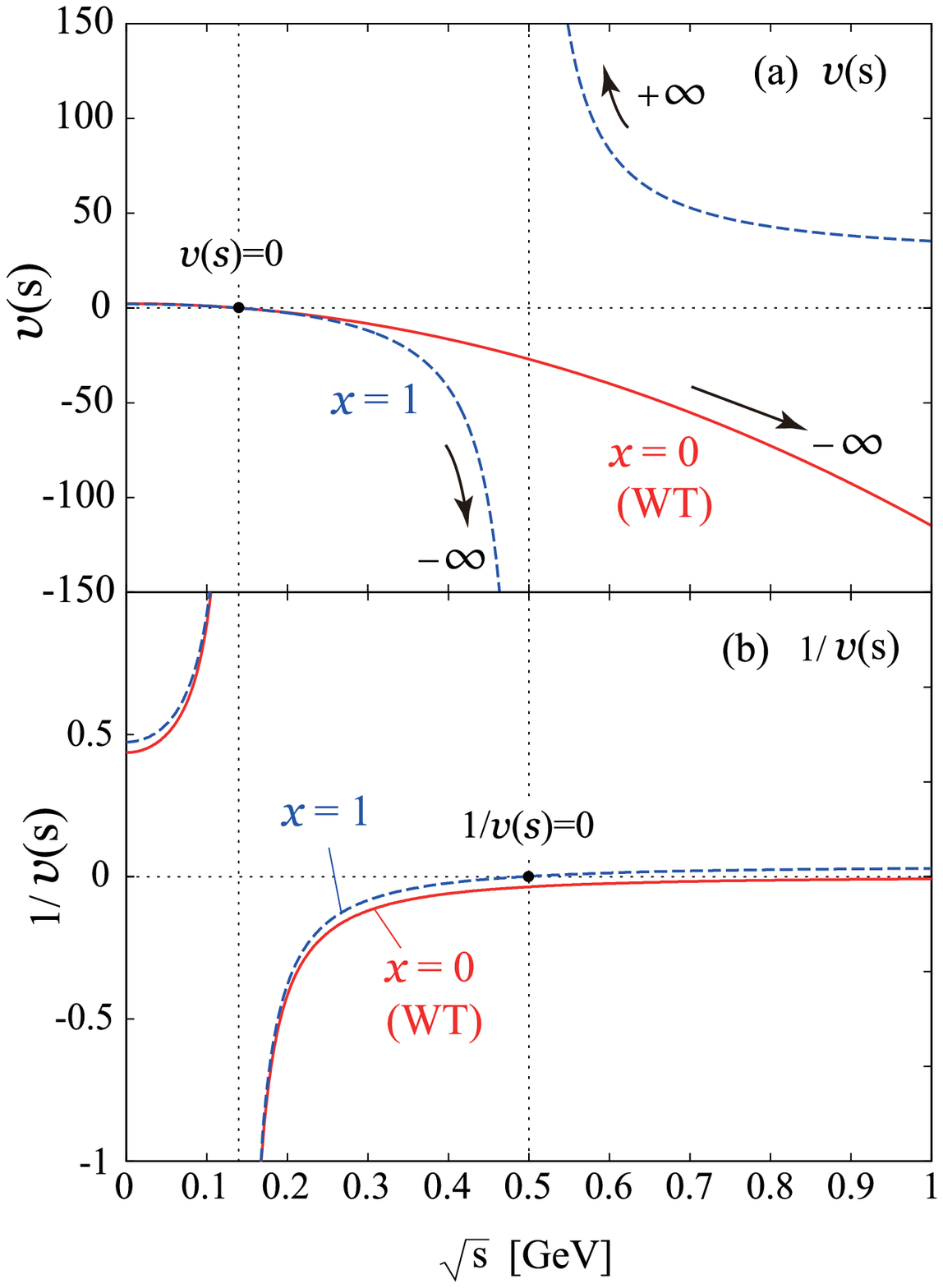}
\caption{(color online) (a) Interaction kernel $v(s)$ in
 Eq.~(\ref{eq:WT+pole_x}) and (b) its inverse $v(s)^{-1}$, with $x=0$
 (the WT term) denoted by the red solid line and $x=1$ (WT + pole) by
 the blue dashed line, as functions of 
 $\sqrt{s}$. The parameters are set to be $m=140$~MeV, $M_0=500$~MeV,
 $f=92.4$~MeV.  The arrows in the figure indicate the divergent
 behaviors of the kernels.
\label{fig:v}
}
\end{figure}

\subsection{Energy independent interaction}
{Now}, we shall revisit {the case of} the constant interaction
\begin{equation}
{v(s)=-c \ (c : \text{positive constant})}
\end{equation}
discussed
in Sec.~\ref{sec:Lurie}, by employing {the} above method to make
the mechanism {of} $Z=0$ clearer.
{To find an}
equivalent Yukawa model,
{we introduce an $s$-dependent $\delta$-term as}
\begin{eqnarray}
 T_\delta(s) &=& \frac{1}{-c + s\delta -G(s)} \nonumber \\
&=& \frac{\frac{1}{\delta}}{s - \frac{1}{c\delta} - \frac{1}{\delta}G(s)} \ .
\end{eqnarray}
The equivalent Yukawa {pole} term is then defined {in} the same way as before as
\begin{equation*}
 V_Y(s)=g_0^2\frac{1}{s-m_\fic^2} \ ,
\end{equation*}
where $m_\fic^2$ and $g_0^2$ are defined as
\begin{equation*}
 m_\fic^2 \equiv \frac{1}{c\delta}, \ \ \ g_0^2 \equiv \frac{1}{\delta} = c m_\fic^2 \ \ .
\end{equation*}

We find again that the {bare} coupling
constant of the Yukawa model is proportional to the mass of the
fictitious {elementary} particle $m_\fic$ 
which diverges in the limit
$\delta\rightarrow 0$
\begin{equation*}
 \lim_{\delta\rightarrow 0} m_\fic^2 = \infty \ , 
 \lim_{\delta\rightarrow 0} g_0^2 = \infty \ , 
\end{equation*}
{and therefore} that the wave function renormalization {constant} becomes
zero $Z=0$ for bound states.  
By employing the above prescription we find that
the mechanisms of $Z=0$ are the same for both
the energy-dependent and energy-independent interactions in the
composite model, where the bare mass of the fictitious elementary
particle should be infinite.
We can see clearly {again} that $Z=0$ is {not} a
consequence of a divergence of the loop function $G$, {nor $G'$}.
This discussion also helps us to distinguish two different
mechanisms {of} $Z=0$ later in Sec.~\ref{sec:BE=0}.


{Here in this section, we have introduced the equivalent elementary
Yukawa model by 
introducing a constant $\delta$ or an $s$-dependent $\delta$-term and
by taking subsequently the limit $\delta\rightarrow 0$.  
In the following section, we will also show that 
the constant shift in $v(s)^{-1}$ causes a divergence in the interaction
$v(s)$ which means a presence of the explicit (elementary) pole term in
the interaction $v(s)$.
It will turn out that the above procedure is closely related to the
regularization scale in the loop function $G$.
}

\section{Multiple interpretations of physical states}
\label{sec:multiple_int}
\subsection{{Cut-off dependence for} one physical state}
{In this section} we would like to discuss an arbitrariness of $Z$ {which} leads to
multiple interpretations of physical states.  To this end,
we consider {a composite model}
with an explicit pole term in addition to the WT interaction such as
\begin{equation}
 v(s)=-\frac{1}{f^2}(s-m^2) + \frac{x}{f^2}(s-m^2)^2\frac{1}{s-M_0^2} \ ,
\label{eq:WT+pole_x}
\end{equation}
{where $x$ is a parameter and $M_0$ the bare mass of an elementary particle.}
This kind of interaction is found in the sigma model {in} the nonlinear
representation~\cite{Oller:1997ti,Hyodo:2010jp,Nagahiro:2013hba,Donoghue:book}.  In general, 
{if the first WT interaction alone can generate a state and the
second}
pole term introduces an 
additional degrees of freedom, the system has two physical
poles~\cite{Nagahiro:2011jn}.  They 
are described as superposition{s} of {the} two basis states; the composite state
developed by the WT term and the elementary particle in the pole term.

Indeed, {by} employing the interaction kernel {$v(s)$} in
Eq.~(\ref{eq:WT+pole_x}),  
we {find} two physical poles with the coefficient $0<x<1$.  {However} one of
them disappears exactly at $x=1$ with
\begin{equation}
 v(s) = - \frac{1}{f^2}(s-m^2) + \frac{1}{f^2}(s-m^2)^2
  \frac{1}{s-M_0^2} \ .
\label{eq:WT+pole}
\end{equation}
The sigma ($\sigma(500)$) resonance in the sigma model {corresponds
to this} case as discussed in detail in Ref.~\cite{Nagahiro:2013hba}.

{Let us} first study {the case} $x=1$ which generates only
one physical pole. 
{In this case the equivalent Yukawa pole term can be obtained {by rewriting}
 Eq.~(\ref{eq:WT+pole}) as,}
\begin{equation}
  v(s) = V_Y(s) = \frac{1}{f^2}(s-m^2)(M_0^2-m^2)\frac{1}{s-M_0^2} \ .
\label{eq:Yukawa_M0}
\end{equation}
{Alternatively, the bare mass of the ``fictitious particle'' $m_\fic^2$ is given by a solution of
Eq.~(\ref{eq:m_fic}) as 
\begin{equation*}
 m_\fic^2=\frac{m^2(M_0^2-m^2)\delta +f^2 M_0^2}{(M_0^2-m^2)\delta+f^2} \ ,
\end{equation*}
which reduces to $M_0^2$ in the limit $\delta\rightarrow 0$ showing that
the equivalent Yukawa pole term~(\ref{eq:V_Y1}) is again given by Eq.~(\ref{eq:Yukawa_M0})}.
{We find that}
the bare coupling $g_0^2=(s-m^2)(M_0^2-m^2)/f^2$ is finite,
{and therefore}
the wave function renormalization {constant} $Z$ {is} non-zero.
{This result is} natural; if there is {an} explicit pole
term as in Eq.~(\ref{eq:WT+pole}), 
$Z$ is finite.  {We can also see
that a different bare mass $M_0$ of the explicit pole term in
Eq.~(\ref{eq:WT+pole}) leads to a different value of $Z$.}


\begin{figure}[thb]
\includegraphics[width=0.4\textwidth]{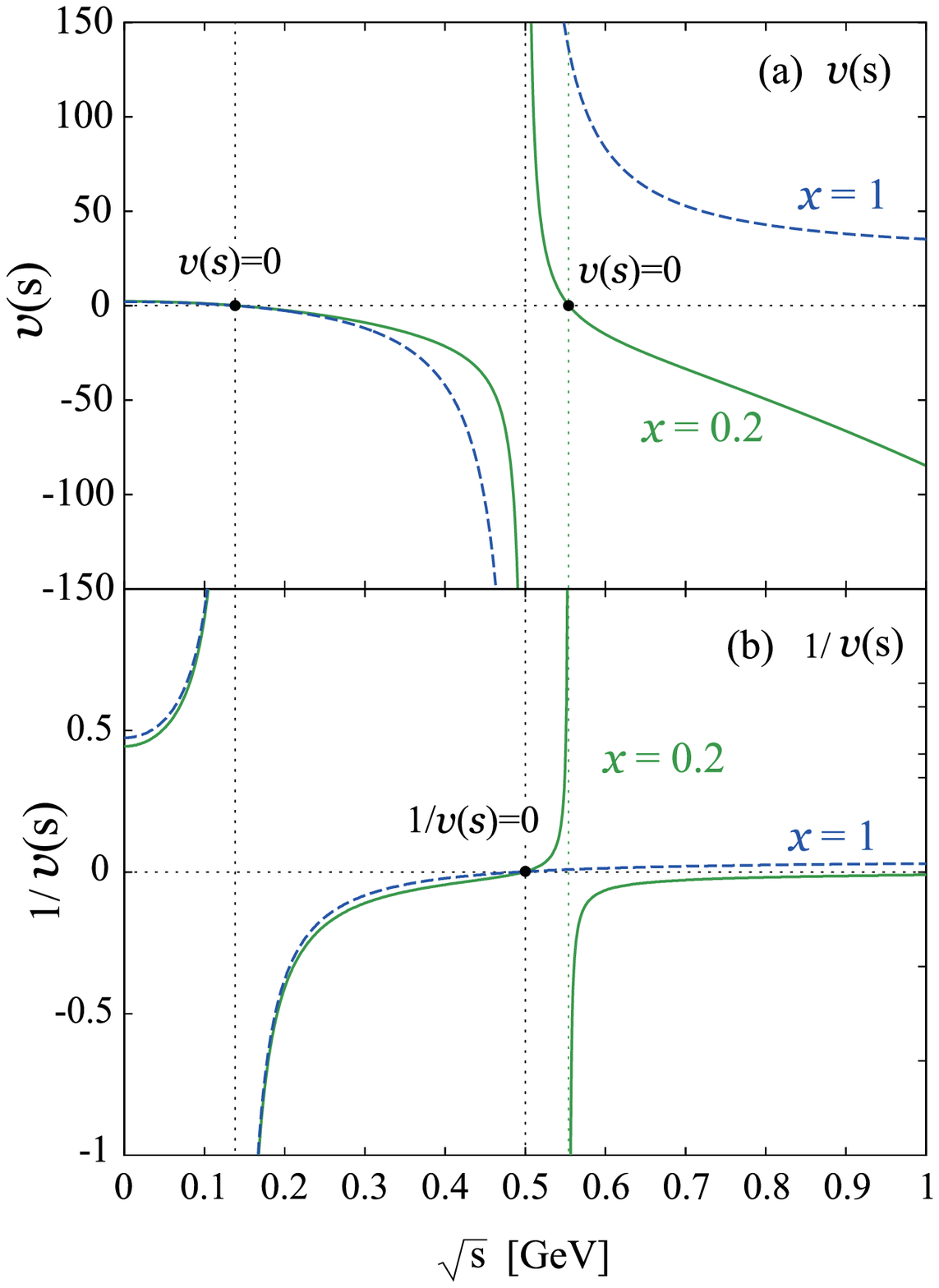}
\caption{(color online) (a) Interaction kernel $v(s)$ in
 Eq.~(\ref{eq:WT+pole_x}) and (b) its inverse $v(s)^{-1}$ with $x=0.2$
 denoted by the green solid line and with $x=1$ by the 
 blue dashed line as functions of $\sqrt{s}$.  
The parameters are set to be $m=140$~MeV, $M_0=500$~MeV, 
 $f=92.4$~MeV.
\label{fig:v2}}
\end{figure}

Now, let us look at the above problem in a different way
{by comparing the two cases of $x=1$ and of $x=0$.}
As shown in
Fig.~\ref{fig:v}, although the shapes of $v(s)$ with ($x=1$) {and} without
($x=0$) the pole term are {quite} different, their inverses $v(s)^{-1}$
{are} quite similar.  Indeed, {they are different} only
{by} a constant $a$, 
\begin{equation}
 v_{x=1}(s)^{-1} = v_{x=0}(s)^{-1} + a \ ,
\end{equation}
{which can be absorbed into} the loop function $G$ {in the
scattering amplitude} as
\begin{eqnarray}
 T(s)^{-1} &=& v_{x=1}(s)^{-1}-G(s) \label{eq:x1G}\\
&=& v_{x=0}(s)^{-1} + a -G(s) \\
&=& v_{x=0}(s)^{-1}-\tilde{G}(s) \label{eq:x0Gtilde}
\end{eqnarray}
where $\tilde{G}(s) \equiv G(s) - a$.  
{This can be done by changing the subtraction constant in the
dimensional regularization {which is equivalent to} 
changing $\Lambda$ in the cut-off 
regularization schemes.  Physically, the change in $G$ corresponds to a
different choice of the dynamical composite model.}
Since the form of the scattering amplitude
(\ref{eq:x0Gtilde}) is the same as discussed in Sec.~\ref{sec:WT}, 
we can conclude that the wave function renormalization {constant} $Z$ is zero for
this system.
The above step is nothing but the one developed in
Ref.~\cite{Hyodo:2008xr}, but we follow it in the opposite way.

Now, {we find that} there are {two} different interpretations
for the physical pole:
\begin{itemize}
 \item[1)] The {physical} pole originates {in} the elementary
	 Yukawa particle with a finite bare mass $M_0$ as in Eq.~(\ref{eq:Yukawa_M0}), {which
	 acquires} the one loop correction, leading {to} $Z\ne 0$.
 \item[2)] The {physical} pole originates in the WT
	 {term} in Eq.~(\ref{eq:WT+pole}) with $Z=0$.
	 {The} pole term in Eq.~(\ref{eq:WT+pole}) {is
	 absorbed} as a counter term {into} the loop {function.}
\end{itemize}
{The above examples show that there is an arbitrariness in the
interpretation.}
We can express the physical state as a pure composite state {with
$Z=0$} as well as an elementary particle with $Z\ne 0$.  In
Eqs.~(\ref{eq:x1G})--(\ref{eq:x0Gtilde}), we have seen 
that the parameter $a$ can be combined either with $v^{-1}$ or with
$G(\tilde{G})$.  Depending on these two schemes, $Z$ can take any
value, while the scattering amplitude $T$ and so physical observables
are invariant.  In other words, $Z$ cannot be determined from
experiments in a model independent manner.}

If we can fix $G$ by using some external condition, we may be able to
{\em partly} avoid such arbitrariness.  This can be done, for instance, by
introducing a cut-off parameter for $G$, {keeping track of its
origin, for instance, to the intrinsic size of the constituents.}

\subsection{Two physical states : two-level problem}

{So far, we have studied the case of $x=1$.  For $0<x<1$, {a} different
feature {appears;} the interaction (\ref{eq:WT+pole_x}) can generate
two physical poles rather than one.}
As shown in Fig.~\ref{fig:v2}, although the interaction kernel $v(s)$
with, say $x=0.2$, {shows} similar {behavior} with $x=1$, the shape of
$v(s)^{-1}$ is quite different. {For $x<1$}, an
additional divergent 
point appears in $v(s)^{-1}$ which {produces} the second physical pole
satisfying $v^{-1}-G=0$.  The difference between $v(s)^{-1}$ with $x=0$
and $x=0.2$, {as can be seen in Fig.~\ref{fig:v2}}, cannot be
absorbed by a constant, nor {by any} 
smooth function.
%
{Therefore, the interpretation 2) {in the previous subsection}
cannot be applied.} 

{Furthermore,
the interpretation 1) also cannot be directly applied. Since we have two
physical states, we cannot {express} the scattering amplitude by
{using} a
single Yukawa pole term $V_Y(s)$.  In fact we
have two solutions of Eq.~(\ref{eq:m_fic}) for $m_\fic$ as
\begin{equation*}
 m_\fic^2=\begin{cases}  
           M_0^2 + {\cal O}(\delta)
	   \xrightarrow{\delta\rightarrow 0} M_0^2\\
	   {\cal O}(\frac{1}{\delta}) \xrightarrow{\delta\rightarrow 0} \infty
	  \end{cases}
\end{equation*}
{indicating}
that there are two ``seeds'' for two physical states.
{The nature of the physical states is different from the previous
ones, leading to the following interpretation;}
\begin{itemize}
 \item[3)] The {physical} pole{s are} mixture{s} of the WT composite
	 state {generated by} the first term in Eq.~(\ref{eq:WT+pole}) and the
	 elementary particle {of} the second term in Eq.~(\ref{eq:WT+pole}).
\end{itemize}
The two physical states are described as
superpositions of {the} two ``seeds'', one is the elementary particle
with the finite mass $M_0$ and the other is the fictitious elementary
particle with infinite bare mass.  This is schematically expressed {as}
\begin{eqnarray*}
 |\text{pole-}a\rangle &=& c_1|m_\fic\rangle + c_2|M_0\rangle \\
 |\text{pole-}b\rangle &=& c_3|m_\fic\rangle + c_4|M_0\rangle  \ ,
\end{eqnarray*} 
which can be analyzed in terms of the two-level
problem~\cite{Nagahiro:2011jn}.}
The component $c_i$ {is} the wave function
renormalization constant $\sqrt{Z}$ {at
each pole position pole-$a$ or pole-$b$}  for {each} basis state .

We note {once again} that {there is arbitrariness in $c_i$'s (or $Z$'s),
depending} on the choice of the basis  
states $|m\rangle$.
For example, as done in Ref.~\cite{Nagahiro:2011jn}, it is possible {to}
first sum up only the WT interaction to obtain the WT composite state,
and redefine the developed pole as ``pure'' composite state
$|m_{WT}\rangle$.  Then we mix it with the
elementary particle, and discuss their mixing,\footnote{If we don't
change the definition of $|M_0\rangle$, then $c_2'=c_2$ and $c_4'=c_4$.}
\begin{eqnarray*}
 |\text{pole-}a\rangle &=& c_1'|m_{WT}\rangle + c_2'|M_0\rangle \\
 |\text{pole-}b\rangle &=& c_3'|m_{WT}\rangle + c_4'|M_0\rangle  \ .
\end{eqnarray*} 
Other definitions of the basis states are {also} possible.
{By choosing} the basis states appropriately, 
{we can discuss} their mixing to understand the
nature of the {physical} states~\cite{Nagahiro:2011jn,Nagahiro:2013hba}.

\subsection{Model dependence of $Z$ : demonstration in the sigma model}
\label{sec:model-dep}
\begin{figure}[thb]
\includegraphics[width=0.3\textwidth]{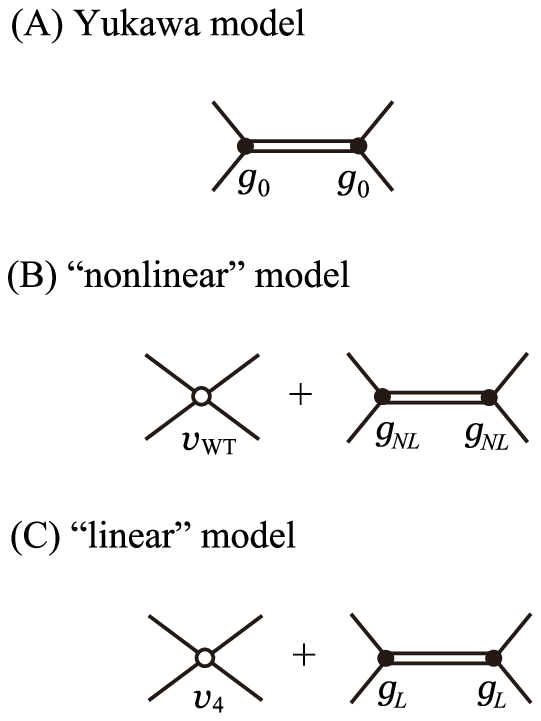}
\caption{Tree amplitudes in (A) {the} Yukawa model {in
 Eq.~(\ref{eq:VY})}, (B) {the} 
 ``nonlinear'' model {in Eq.~(\ref{eq:VNL})}, 
 and (C) {the} ``linear'' model {in Eq.~(\ref{eq:VL})}.  {The solid and
 open circles are for the three-point vertices and for the four-point
 interactions, respectively.}
\label{fig:trees}}
\end{figure}

Finally in this section, we discuss further 
a model dependence of $Z$.
In Sec.~\ref{sec:compositenessZ}, we have introduced the Yukawa model {as}
\begin{eqnarray}
V_Y(s) &=& \frac{(s-m^2)(m_\fic^2-m^2)}{f^2}\frac{1}{s-m_\fic^2} 
 \nonumber \\
&\equiv& g_0^2(s)\frac{1}{s-m_\fic^2} \ ,
\label{eq:VY}
\end{eqnarray}
which has only the three-point interaction and is equivalent to the
composite model with the WT interaction.  
{This expression can be further rewritten into two different forms as}
\begin{eqnarray}
 V_{NL}(s)&=&-\frac{(s-m^2)}{f^2} +
  \frac{(s-m^2)^2}{f^2}\frac{1}{s-m_\fic^2}  \nonumber \\
&\equiv& v_{WT}(s)+g_{NL}^2(s)\frac{1}{s-m_\fic^2}, \label{eq:VNL}
\end{eqnarray}
and
\begin{eqnarray}
 V_{L}(s)&=&\frac{(m_\fic^2-m^2)}{f^2} +
  \frac{(m_\fic^2-m^2)^2}{f^2}\frac{1}{s-m_\fic^2}  \nonumber \\
&\equiv& v_{4}+g_L^2\frac{1}{s-m_\fic^2} \ ,
\label{eq:VL}
\end{eqnarray}
{which might define two different ``models'' (or more concretely
different diagrams) as
depicted Fig.~\ref{fig:trees}.}
{Here we refer to the second model as ``nonlinear (NL) model'' and
the third as ``linear (L) model'' for sake of simplicity\footnote{The
{interactions} $V_{NL}(s)$ and $V_{L}(s)$ correspond to  
the tree amplitudes of $\pi\pi$ scattering {in} the nonlinear and linear
representations {of} the sigma model used in
Refs.~\cite{Hyodo:2010jp,Nagahiro:2013hba}.}.} 
{The interaction}
 $V_{NL}(s)$ {in Eq.~(\ref{eq:VNL})} consists of the WT term
{$v_{WT}(s)$} and the pole term {with the {energy-dependent bare}
coupling $g_{NL}(s)$}.  {The interaction}
$V_L(s)$ {in
Eq.~(\ref{eq:VL})} has
a repulsive {four-point} interaction {$v_4$}
and the pole term with the constant coupling {$g_L$}.

Although these {interactions} contain {the same propagator
$(s-m_\fic^2)^{-1}$ of} a fictitious elementary particle
{and the resulting scattering 
amplitude are the same}, the corresponding wave
function renormalization {constants} for these {fictitious} particles are different.

{For example}, in the linear {model} the wave function renormalization
{constant} $Z_{{L}}$ in
the limit $\delta\rightarrow 0$ $(m_\fic\rightarrow\infty)$ is not zero,
while in the nonlinear {or in the} Yukawa {model} it is zero, {$Z_{NL}=Z_Y=0$}. 
{In the linear model, the scattering amplitude is given by
\begin{equation*}
 T(s)=T_{4}(s)+{\tilde{g}_{L}}^2(s) \Delta_{L}(s) \ ,
\end{equation*}
where $T_{4}$ is defined by
\begin{equation*}
 T_{4}(s)=\frac{1}{v_{4}^{-1}-G(s)} \ ,
\end{equation*}
with the repulsive four-point interaction $v_4$~\cite{Weinberg:1962hj}.
The coupling ${\tilde{g}_{L}}$ contains vertex corrections due to the
contact interaction $v_4$ as
\begin{equation*}
 {\tilde{g}_{L}}(s) = \frac{g_{L}}{1-v_{4}G(s)} \ .
\end{equation*}
The dressed propagator $\Delta_{L}$ is given by
\begin{equation*}
 \Delta_{L}(s) = \frac{1}{s-m_\fic^2-\Pi_{L}(s)} \ ,
\end{equation*}
with the self-energy 
\begin{eqnarray}
 \Pi_{L}(s) &=&
  g_{L}^2\frac{G(s)}{1-v_{4}G(s)} \nonumber \\
&=&\frac{(m_\fic^2-m^2)^2}{f^2}\frac{G(s)}{1-\frac{(m_\fic^2-m^2)}{f^2}G(s)}
  \ ,
\label{eq:PiL}
\end{eqnarray}
\begin{figure}[hbt]
\includegraphics[width=0.4\textwidth]{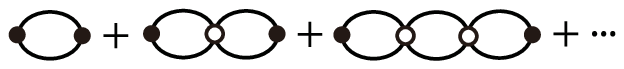}
\caption{Self-Energy for the fictitious particle in the linear and {nonlinear}
 models.   {The} solid and open {circles} indicate the couplings in the same
 manner {as in} Fig.~\ref{fig:trees}. 
\label{fig:PiL}}
\end{figure}
dipicted in Fig.~\ref{fig:PiL}.  The wave function renormalization
{constant} $Z_{L}$ for the elementary particle is defined by using the
self-energy as,}
\begin{equation*}
 Z_{L} =
  \left(1-\left.\Pi_{L}'(s)\right|_{s=\mu_\delta^2}\right)^{-1} \ .
\end{equation*}
The derivative $\Pi_L'$ is calculated as
\begin{equation}
 \left.\Pi_L'(s)\right|_{s=\mu^2_\delta} =
  \frac{1}{f^2}\frac{G'(\mu_\delta^2)}{\left(
\frac{1}{m_\fic^2-m^2}-\frac{1}{f^2}G(\mu_\delta^2)\right)^2}
\label{eq:Pi_L1}
\end{equation}
and in the limit $\delta\rightarrow 0$ it reduces to
\begin{equation}
 \left.\Pi_L'(s)\right|_{s=\mu^2_\delta}
 \xrightarrow[m_\fic\rightarrow\infty]{\delta\rightarrow 0}
f^2\frac{G'(\mu^2)}{G(\mu^2)^2} \ .
\label{eq:Pi_L2}
\end{equation} 
If $G(s)$ and $G'(s)$ are regularized in the same manner as in the
previous section, this derivative takes a finite value and then $Z_L$ is  
not equal to zero even in the limit $\delta\rightarrow 0$.

{In the nonlinear {model} the scattering amplitude and the dressed
propagator can be derived in a similar manner.  The self-energy $\Pi_{NL}$ is
now given by}
\begin{equation}
 \Pi_{NL}(s)=\frac{(s-m^2)^2}{f^2}\frac{G(s)}{1+\frac{(s-m^2)}{f^2}G(s)}  .
\label{eq:PiNL}
\end{equation}
The derivative of the self-energy is obtained by
\begin{eqnarray}
 \left.\Pi_{NL}'(s)\right|_{s=\mu^2_\delta} &=&
 \frac{2G(\mu_\delta^2)}{\delta} +
 \frac{f^2G'(\mu_\delta^2)-G(\mu_\delta^2)^2}{\delta^2} \ .
\label{eq:Pi_L}
\end{eqnarray} 
{Obviously, this quantity}
diverges in the limit 
$\delta\rightarrow 0$ {leading to} $Z_{NL}\rightarrow 0$.  In the
nonlinear case $\lim_{\delta\rightarrow 0}Z_{NL}=0$ can be explained also by
using the two-level problem as discussed in Ref.~\cite{Nagahiro:2013hba}.

\begin{figure*}[tbt]
\includegraphics[width=0.7\textwidth]{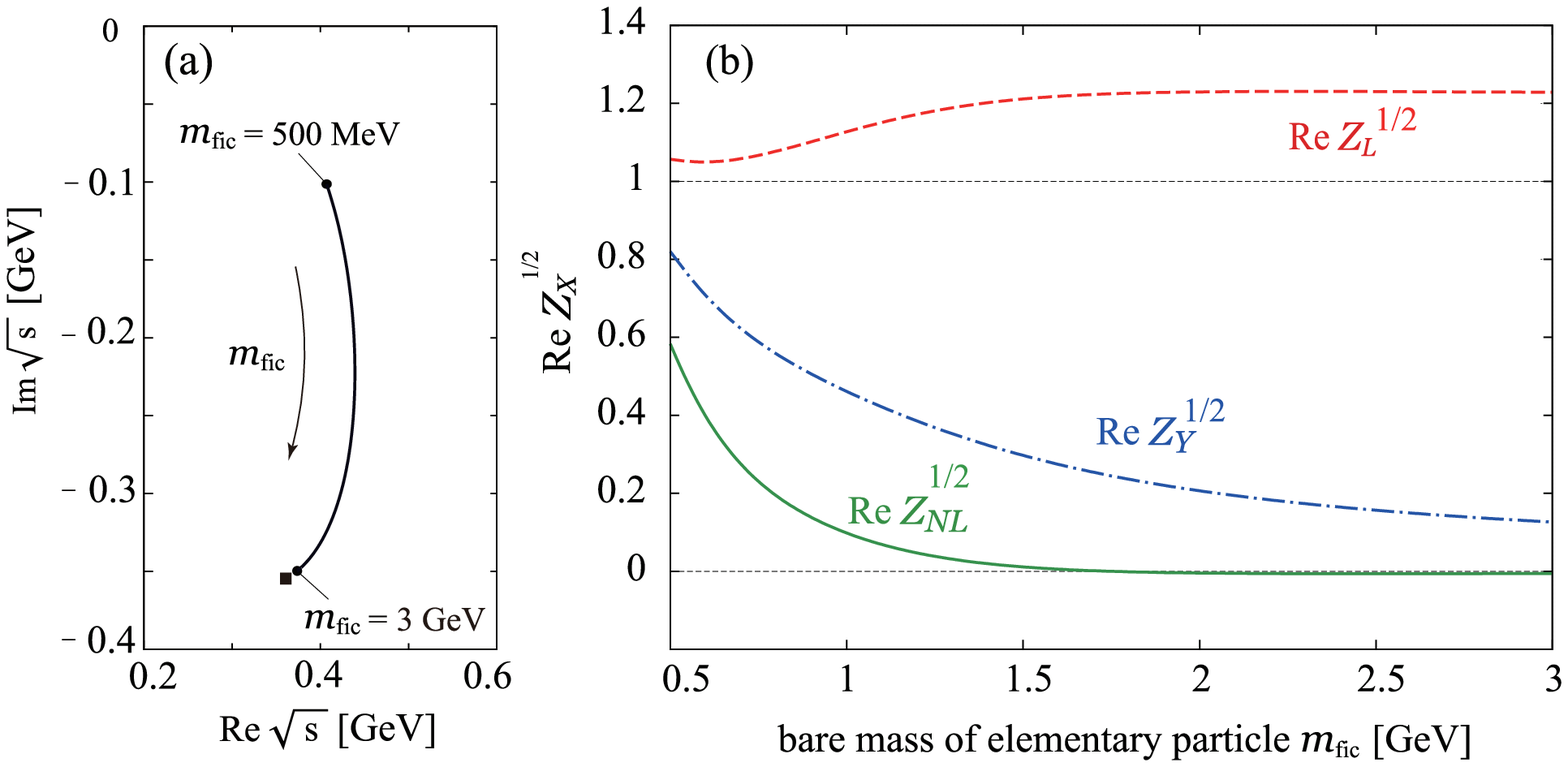}
\caption{(color online) (a) Flow of the pole position of the sigma ($\sigma(500)$) resonance
 in the scattering amplitude obtaining by changing the bare
 mass of the elementary particle (sigma meson) from 0.5~GeV to 3~GeV.
 The solid square indicates the pole position in the limit
 $m_\fic\rightarrow \infty$~\cite{Nagahiro:2013hba}.   (b) The real part
 of the wave function 
 renormalization constants as functions of the bare mass of the
 elementary particle.
The blue dash-dotted line is for the Yukawa model $(Z^{1/2}_Y)$ with the
 interaction kernel (\ref{eq:vy}), the green solid line for 
the nonlinear model $(Z^{1/2}_{NL})$ with
 (\ref{eq:vnl}), and the red dashed line for the linear model
 $(Z^{1/2}_L)$  with (\ref{eq:vl}).
\label{fig:Z3_ms}}
\end{figure*}

Here we demonstrate the model dependence of 
$Z$ by using the sigma ($\sigma(500)$) meson in the sigma model.  {In
Fig.~\ref{fig:Z3_ms}, we show the wave function 
renormalization {constant} $Z$ for the three cases; with the Yukawa model
{($Z_Y$)}, the nonlinear model {($Z_{NL}$)} and the linear model
{($Z_L$)}.  The parameter $m$ is set to be {the mass of the pion}
$m=m_\pi=138$~MeV, $f$ the pion decay constant $f=f_\pi=92.4$~MeV.  The
mass $m_\fic$ corresponding to the bare mass of the elementary sigma meson
is varied from 0.5~GeV to 3~GeV in the figure.} 

{Here we note that
 for the isosinglet sigma meson we employ $3V_X(s)+V_X(t)+V_X(u)$
 instead of $V_X(s)$ alone as the interaction kernel $(X=L$, $NL$, and $Y$).
Although {the inclusion of} the $t$ and $u$ channels changes the expressions of
the self-energies in Eqs.~(\ref{eq:PiY}), (\ref{eq:PiL}) and
(\ref{eq:PiNL}), the conclusion about the fate of $Z_X$ for the large
$m_\fic$ limit {is} not affected.
The $s$-wave
tree amplitude for the $s$-wave resonance can be projected out {by},
\begin{equation}
 v(s)=\frac{1}{2}\int_{-1}^1 dx T_{I=0}^{\rm tree}(s,t(x),u(x))
  P_{\ell=0}(x)
\label{eq:swave}
\end{equation}
in the center of mass frame.  The result of the projection is given by
\begin{eqnarray}
v(s)_Y&=&g_0^2(s)\frac{1}{s-m_\fic^2} + v^{\rm ex}_Y(s)  \label{eq:vy}\\
v(s)_{NL}&=&g_{NL}^2(s)\frac{1}{s-m_\fic^2} + v^{\rm ex}_{NL}(s) + v^{\rm con}_{NL}(s) \label{eq:vnl} \\
v(s)_{L}&=&g_{L}^2\frac{1}{s-m_\fic^2} + v^{\rm ex}_{L}(s) + v^{\rm
 con}_{L}(s) \ , \label{eq:vl}
\end{eqnarray}
where the bare couplings are now defined by
\begin{eqnarray}
 g_0^2(s) &=& \frac{3}{f^2}(s-m^2)(m_\fic^2-m^2) \label{eq:3g0}\\
 g_{NL}^2(s) &=& \frac{3}{f^2}(s-m^2)^2 \label{eq:3gNL}\\
 g_{L}^2 &=& \frac{3}{f^2}(m_\fic^2-m^2)^2 \label{eq:3gL}\ ,
\end{eqnarray}
and $v^{\rm ex}$ and $v^{\rm con}$ denote the
($s$-wave projected) $t$- and $u$-channel exchange and the contact terms, respectively.
The concrete forms of (\ref{eq:vy})--(\ref{eq:vl}) are summarized in
Appendix \ref{sec:App}.}

Since the {three} {interaction kernels}
{(\ref{eq:vy})--(\ref{eq:vl})} are the same, the pole position of the
physical $\sigma(500)$ resonance {is} the same {for} all cases, as shown in
Fig.~\ref{fig:Z3_ms}(a)~\cite{Nagahiro:2013hba}. However, the wave
function renormalization {constant}
$Z$ for the three cases are different as shown in
Fig.~\ref{fig:Z3_ms}(b). 
{In the} Yukawa {model} $Z_Y$ decreases and approaches zero as $m_\fic$ is
increased as discussed before.  {In} the nonlinear {model} $Z_{NL}$ also
decreases, and finally {becomes} zero in the limit
$m_\fic\rightarrow \infty$.\footnote{The nonlinear $Z_{NL}^{1/2}$ is
the same as $z_{22}^{1/2}$ in Fig.~10(b) of Ref.~\cite{Nagahiro:2013hba}.}
{In contrast,}
we find that the $Z_L$ {of the linear model} shows quite different behavior from
the others.  It {remains} finite even in the large $m_\fic$ limit.
These are consistent with the {discussion in Eq.~(\ref{eq:Pi_L1})
and afterwards.} 

{If} we {use} the finite value of $Z_L$ in the linear
model, {we may interpret that} the
$\sigma(500)$ resonance {has} a large component of the elementary
particle even if its bare mass is infinite.  It does not conflict,
however, to the other interpretations in the Yukawa and nonlinear models
in which the elementary component is zero with the infinite bare mass,
because each value of $Z$ indicates the probability of finding {the}
elementary particle {defined} in {\em each model}:
{the elementary particles in different models are different.}

\section{Zero energy bound system}
\label{sec:BE=0}
{In this section} we discuss the zero-energy bound state
which also leads exactly to $Z=0$, but the underlying mechanism {is}
very much different from what we have discussed so far.
{Let us {recall} that the mechanism of $Z=0$}
discussed in the {previous sections, where}
\begin{equation}
 g_0\rightarrow \infty\ (\text{then}\ m_\fic\rightarrow \infty) \ \ \& \ \
  g_R = \text{finite}  
\label{eq:mech1}
\end{equation}
{that is}
the bare mass of the fictitious particle must be infinite
{and far away from the energy region of interest.}
{In contrast, there is another mechanism which also leads to $Z=0$ as}
\begin{equation}
 g_0 = \text{finite}\ (\text{then}\ m_\fic = \text{finite)} \ \& \ \ g_R
  = 0 \ .
\label{eq:mech2}
\end{equation}
{This can be realized when the physical state appears at the
threshold, namely zero-energy bound state.}
{In fact, it was shown that} the renormalized coupling constant
$g_R^2$ is proportional to $\sqrt{|B|}$
{for small $|B|$~\cite{Nambu:1961zz,Weinberg:1962hj},}
\begin{equation*}
 g_R^2 \propto \sqrt{|B|} \ ,
\end{equation*}
with the binding energy $B$.
{Consequently,} $Z\rightarrow 0$ in the limit $B=0$.  
{The behavior of $g_R$ at the threshold can be directly seen from
Eq.~(\ref{eq:GR}) for the constant interaction {case} and from
Eq.~(\ref{eq:gR1}) for the WT {interaction} {case},  {where we can see that} $g_R=0$ due to the
divergence of $G'(s)$ at the threshold~\cite{PhysRevC.85.015201}.}  

\begin{figure}[hbt]
\includegraphics[width=0.4\textwidth]{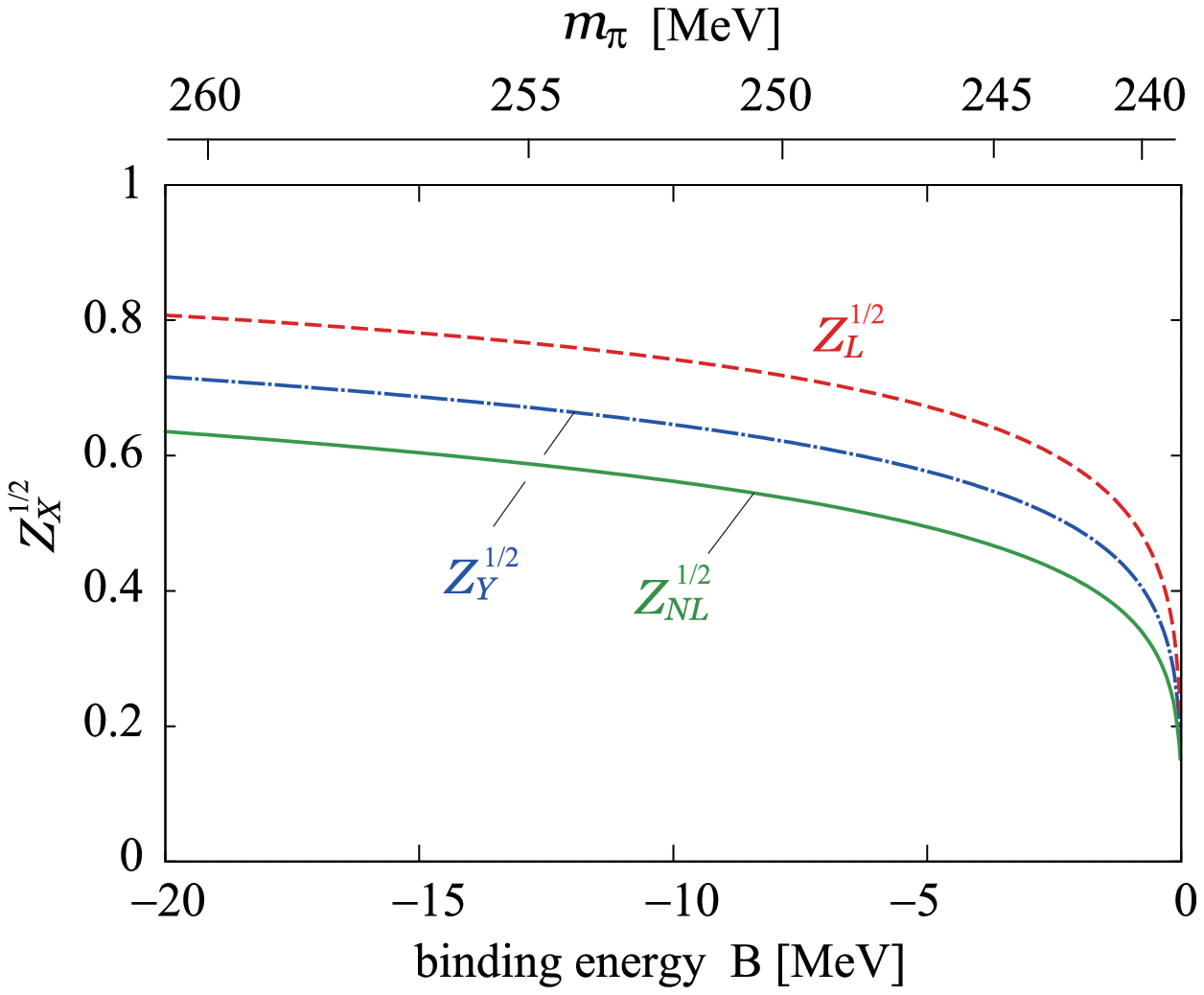}
\caption{(color online) Wave function renormalization constants as
 functions of the binding energy.
The blue dash-dotted line is for the Yukawa model ($Z_Y^{1/2}$)
 with the interaction kernel (\ref{eq:vy}), the green solid line for 
the nonlinear 
 model ($Z_{NL}^{1/2}$) with (\ref{eq:vnl}), and the red dashed line for 
 the linear model ($Z_{L}^{1/2}$) with (\ref{eq:vl}).
The parameters are set to be $f=f_\pi=92.4$ and $m_\fic=550$~MeV.  The
 corresponding pion mass is indicated in the upper horizontal axis.
\label{fig:BE_Z0}
}
\end{figure}

{The important point here is that,
the condition $Z=0$ for the zero-energy bound state has no relation to
the bare mass of the elementary particle and does not mean {its} infinite
{value}.  As an example,}
{in Fig.~\ref{fig:BE_Z0}, we show the wave function renormalization
{constants} for
bound states
in the three representations; the Yukawa ($Z_Y$), nonlinear
($Z_{NL}$), and linear ($Z_L$) models, {as functions of the binding
energy {$B = \mu-2m_\pi$  where $\mu$ is the mass of the bound state.}}  The models used here are the same
as for Fig.~\ref{fig:Z3_ms} except the mass of the pion {which is set lager} 
to obtain a bound state {with a finite $B$} {below the threshold} {as indicated in Fig.~\ref{fig:BE_Z0}}.
The bare mass of the elementary sigma meson is fixed to be
$m_\fic=550$~MeV.}
As discussed in Sec.~\ref{sec:model-dep}, the three wave function
renormalization constant{s} $Z_X$ are different {from} each other at finite binding energies
$B$, but becomes zero at $B=0$ for all cases.  
{However, 
$Z=0$ for the zero-energy bound state does not mean that the elementary particle is
irrelevant in this system.  Indeed, 
the energy (mass) of the physical state ($\sim 480$~MeV in this case) is
determined by the bare mass of the elementary particle (550~MeV)
together with the self-energy with {\em finite} $g_0$ and {\em finite} (regularized)
$G$, which is independent from the divergence of $G'$.}
Such a situation can arise in any composite state close to the threshold.

The ``compositeness'' {or ``elementarity''} is often discussed
without distinction between 
these two mechanisms (\ref{eq:mech1}) and (\ref{eq:mech2}).  {Although both gives the
divergence of the derivative of the self-energy $\Pi'=(g_0^2G)'$}, they
should be discussed separately 
because the physical meanings {are} different.  The so-called
``{elementarity}'' we would like to know is the former one, that is
to say whether the energy (mass) of the elementary particle is
far from the physical pole or not.  
The condition $Z=0$ for the zero-energy bound state does not {exclude} the
existence of the elementary state close to the physical state.

\section{{Summaries and discussions}}
\label{sec:conclusion}
We have discussed the ``compositeness'' or ``elementarity'' of
{composite} state{s} {by means of} the wave function
renormalization {constant} {$Z$}.   
We have shown that {$s$-wave scattering amplitudes and} $s$-wave
state{s} generated {dynamically} by an 
energy-dependent {or independent} interaction can be
equivalently represented by a Yukawa model
{with an energy-dependent or
independent coupling and with the infinite bare mass of a fictitious elementary particle.}
{Consequently,}
the wave function renormalization {constant} $Z$ for any composite state
{can be} zero, which means {that}
the corresponding bare elementary field $\phi=Z\phi_R$
vanishes.  
The idea can be equally {applied to both} resonant and bound
states.
Here the underlying mechanism {of} $Z=0$ is that the bare coupling
and the bare mass of the corresponding Yukawa particle 
{{become} infinite to be consistent {with} the composite state.}

We have also discussed the case {of the zero energy bound state,}
which also leads to $Z=0$.  
This is due to the divergence of the derivative of the loop function
at the threshold, {which should be distinguished from the above
mentioned mechanism of}
the infinite bare mass of the fictitious elementary particle.
The condition $Z=0$ for the zero-energy bound
state does not exclude an elementary state near the physical
state.  
{Therefore we should be careful when we discuss the ``elementarity'' of
barely bound systems.}

{The argument for the condition $Z=0$ with infinite
bare mass corresponds to the assertion made by Weinberg in
Ref.~\cite{Weinberg:1962hj} that any physical state can be equivalently 
{represented}
by a ``quasi-particle'' with infinite bare mass and hence with $Z=0$.
{To see what this statement means more clearly, let us consider hadron
resonances in the chiral unitary
approach~\cite{Jido:2003cb,PhysRevC.65.035204}.
There,} it is widely believed that  
$\Lambda(1405)$ is a good candidate of a composite state of weakly bound $\bar K N$ 
molecular state, while $N(1535)$ contains to a large
extent non-composite component~\cite{Hyodo:2008xr}.
However, {according to} the assertion, both particles can be always made
``composite'' with zero elementarity, $Z=0$. }

{At first sight this statement sounds inconsistent.  
However, a solution can be given 
if we look at the setup of the chiral unitary approach
which is defined together with a natural cut-off scale 
corresponding to {the intrinsic size of}
the constituent hadrons.  
{While}
the properties of $\Lambda(1405)$ can be well reproduced 
within the natural framework of the model, those
of $N(1535)$ {can be so} when a cut-off scale 
is chosen at a value which is different from the natural value.  
As we will explain shortly, the use of the un-natural cut-off 
scale introduces a pole term in the interaction.  
We can then measure the compositeness by means of the 
elementary particle corresponding to the pole term, which leads to 
a finite value $Z \neq 0$.
Without the pole term, as we have discussed so far, the renormalization
constant $Z$ is equal to zero.
In view of this, the value of the renormalization constant itself does
not tell us the nature of the {physical state}.}

{Let us now look at the problem in a slightly different manner in
terms of the physical observables, that is the scattering amplitude.
Suppose we}
determine the cut-off {value} $\Lambda=\Lambda_h$ at
the hadronic scale.
However, the resulting scattering amplitude does not always reproduce observables.  
Then we attempt to change $\Lambda$ from $\Lambda_h$ to reproduce the
observables.
In the scattering amplitude, this amounts to the change in the loop function, 
$G(\Lambda_h) \rightarrow G(\Lambda)$.  
The important point that should be emphasized here is that the difference 
$G(\Lambda) - G(\Lambda_h)$ introduces the new pole interaction as, 
$v(s) \rightarrow \tilde{v}(s) = v(s)$ + pole term~\cite{Hyodo:2008xr}.  
Then we can evaluate the renormalization constant $Z$ in terms of the
elementary particle  
associated with the new pole term~\cite{Nagahiro:2013hba}.
{In this manner}, the resulting $Z$ can take an arbitrary finite
value.
In other words, while the physical observable {(amplitude)} is invariant
under the simultaneous changes in $G$ and the interaction $v$, the
renormalization constant $Z$ cannot 
be determined from experiment in a model independent manner. 
{This is an unavoidable feature because the renormalization constant
$Z$ is not a physical observable.}

To discuss the ``elementarity'', we need to {first specify} a reasonable (or
useful) framework for the dynamics of the system to make a 
proper description for hadrons.
This is to a large extent a question of economization.
{The criterion for choosing a suitable model should be given by
external conditions independently from the present discussions
concerning the constant $Z$.}

\section*{Acknowledgement}
The authors are grateful to T. Hyodo and T. Sekihara for various
discussions. This work is supported by Grants-in-Aid for Scientific
Research (Nos.~24105707 and 26400275 for H.~N.)  and (Nos. 21105006(E01)
and 26400273 for A.~H.).

\appendix

\section{tree amplitude for the $\pi\pi$ scattering}
\label{sec:App}
Here we show the concrete form of the interaction kernels in the
$\pi\pi$ scattering amplitude in the sigma model~\cite{Donoghue:book,Oller:1997ti,Hyodo:2010jp,Nagahiro:2013hba}.
The tree amplitude for the isospin $I=0$ channel is determined in terms
of a sigle function $V(s,t,u)$ as
\begin{equation*}
 T_{\rm tree} = 3V(s,t,u) + V(t,s,u)+V(u,t,s).
\end{equation*}
The function $V(s,t,u)=V(s)$ is given by $V_{NL}$ in Eq.~(\ref{eq:VNL})
in the nonlinear model~\cite{Oller:1997ti,Hyodo:2010jp}, 
$V_{L}$ in Eq.~(\ref{eq:VL}) in the linear model~\cite{Hyodo:2010jp},
and $V_Y(s)$ in Eq.~(\ref{eq:VY}) in the Yukawa model.
The
$s$-wave projection is performed by Eq.~(\ref{eq:swave}) and the results
of the projection are given by,
\begin{widetext}
\begin{eqnarray}
v(s)_Y&=&g_0^2(s)\frac{1}{s-m_\fic^2} -
\frac{2}{f^2}(m_\fic^2-m^2)+\frac{1}{f^2}\frac{2(m^2-m_\fic^2)^2}{s-4m^2}\ln\left(\frac{m_\fic^2}{m_\fic^2-4m^2+s}\right) \\
v(s)_{NL}&=&g_{NL}^2(s)\frac{1}{s-m_\fic^2} -
  \frac{1}{f^2}(2s-m^2)+\frac{1}{f^2}\left\{-(s-2m_\fic^2)+
\frac{2(m^2-m_\fic^2)^2}{s-4m^2}\ln\left(\frac{m_\fic^2}{m_\fic^2-4m^2+s}\right)
\right\} \\
v(s)_L&=&g_L^2\frac{1}{s-m_\fic^2} 
+\frac{5}{f^2}(m_\fic^2-m^2)+\frac{1}{f^2}\frac{2(m^2-m_\fic^2)^2}{s-4m^2}\ln\left(\frac{m_\fic^2}{m_\fic^2-4m^2+s}\right)
\end{eqnarray}
\end{widetext}
where the coupling $g_0^2$, $g_{NL}^2$, and $g_L^2$ are defined by Eqs.~(\ref{eq:3g0})--(\ref{eq:3gL}).

\bibliography{a1_nature.bib}
\end{document}